\documentclass[onecollarge,natbib]{svjour2}

\bibpunct{[}{]}{,}{n}{}{,} 
\smartqed  
\usepackage{graphicx}
\usepackage{subfig}

\usepackage{mathptmx}   
\usepackage{multirow}
\usepackage{amsmath}
\usepackage{comment}

\usepackage{lineno}

\newcommand{\beq}{\begin{eqnarray}}
\newcommand{\eeq}{\end{eqnarray}}
\newcommand{\be}{\begin{eqnarray*}}
\newcommand{\ee}{\end{eqnarray*}}

\newcommand{\vect}[1]{\vec{#1}}

\newcommand{\ie}{{\it i.e.}}
\newcommand{\eg}{{\it e.g.}}

\newcommand{\ct}[1]{{Table.~\ref{#1}}}

\def\lsim{\raise0.3ex\hbox{$<$\kern-0.75em\raise-1.1ex\hbox{$\sim$}}}
\def\gsim{\raise0.3ex\hbox{$>$\kern-0.75em\raise-1.1ex\hbox{$\sim$}}}

\def\pp   {$pp$}

\def\jpsi   {\mbox{$J/\psi$}}

\def\beq     {\begin{equation}}
\def\eeq     {\end{equation}}
\def\upsi    {\mbox{$\Upsilon$}}

\def\pp   {$pp$}

\def\jpsi    {\mbox{$J/\psi$}}

\def\beq     {\begin{equation}}
\def\eeq     {\end{equation}}
\newcommand{\sqrtS}[1]{\mbox{$\sqrt{s_{#1}}$}}
\newcommand{\psip}{\mbox{$\psi'$}}
\newcommand{\chic}{\mbox{$\chi_c$}}
\newcommand{\chib}{\mbox{$\chi_b$}}

\usepackage{ifpdf}
\ifpdf
\usepackage[pdftex]{hyperref}
\else
\usepackage[hypertex]{hyperref}
\fi

\hypersetup{
  pdftitle={Quarkonium Physics at a Fixed-Target Experiment using the LHC Beams},%
  pdfauthor={J.P. Lansberg, S.J. Brodsky, F. Fleuret, C. Hadjidakis},%
  pdfsubject={},%
  pdfkeywords={Quarkonium Production, Fixed-Target experiment, Large Hadron Collider},%
  pdfstartview={},%
  bookmarksopen=true, breaklinks=true, debug=true, %
  colorlinks=true, linkcolor=red, citecolor=blue, urlcolor=blue
}

\journalname{Few Body Systems}
\begin{document}

\title{Quarkonium Physics at a Fixed-Target Experiment using the LHC Beams\thanks{Presented at the workshop ``30 years of strong interactions'', Spa, Belgium, 6-8 April 2011.}}

\author{J.P. Lansberg \and S.J. Brodsky \and F. Fleuret \and C. Hadjidakis}%

\institute{  
J.P. Lansberg, C. Hadjidakis \at IPNO, Universit\'e Paris-Sud, CNRS/IN2P3, 91406 Orsay, France
\and 
S.J. Brodsky \at SLAC National Accelerator Laboratory, Theoretical Physics, Stanford University, Menlo Park, California 94025, USA 
\and 
F. Fleuret \at  Laboratoire Leprince Ringuet, Ecole polytechnique, CNRS/IN2P3, 91128 Palaiseau, France\\
}

\date{Received: date / Accepted: date}

\maketitle

\begin{abstract}
We outline the many quarkonium-physics opportunities offered by a multi-purpose fixed-target experiment
using the $p$ and Pb LHC beams extracted by a bent crystal. This provides an integrated luminosity 
of 0.5 fb$^{-1}$ per year on a typical 1cm-long target. Such an extraction mode
does not alter the performance of the collider experiments at the LHC.
With such a high luminosity, one can analyse quarkonium production in great details 
in $pp$, $pd$ and $pA$ collisions  at $\sqrt{s_{NN}} \simeq 115$ GeV  and at $\sqrt{s_{NN}}\simeq 72$ GeV in Pb$A$ collisions. 
In a typical $pp$ ($pA$) run, the obtained quarkonium yields per unit of rapidity are  
2-3 orders of magnitude larger than those expected at RHIC and about respectively 10 (70) times larger
than for ALICE. In Pb$A$, they are comparable. 
By instrumenting the target-rapidity region, the large negative-$x_F$ domain can be accessed for the first time,
greatly extending previous measurements by Hera-B and E866. Such analyses should help resolving the quarkonium-production
 controversies and clear the way for gluon PDF extraction via quarkonium studies. 
The nuclear target-species versatility provides a unique opportunity 
to study nuclear matter and the features of the hot and dense matter formed in Pb$A$ collisions.
A polarised proton target allows the study of transverse-spin asymmetries in 
$J/\psi$ and $\Upsilon$ production, providing access to the gluon and charm Sivers functions.
\end{abstract}

\tableofcontents

\section{Introduction}
\label{intro}

Fixed-target hadroproduction experiments have played a major role in quarkonium physics, beginning with
the co-discovery of the $J/\psi$~\cite{Aubert:1974js} at BNL in 1974,  the discovery of 
the $\Upsilon$~\cite{Herb:1977ek} and the first observation of $h_c$~\cite{Armstrong:1992ae} at Fermilab.
Fixed-target experiments have revealed many novel and unexpected features of quark and gluon
dynamics, including the anomalous suppression of $J/\psi$~\cite{Abreu:2000ni} 
in PbPb collisions at SPS, the strong non-factorising nuclear suppression of $J/\psi$ hadroproduction at high 
$x_F$~\cite{Hoyer:1990us} and the large-$x_F$ production of $J/\psi$ pairs~\cite{Badier:1982ae}.

As we have discussed in Ref.~\cite{Brodsky:2012vg}, the collisions of the LHC proton and heavy-ion 
beams on fixed targets provide a remarkably large range of physics opportunities thanks to the high 
luminosity typical of fixed-target experiment and due to
its relatively high center-of-mass (CM) energy, 115 GeV per nucleon with the proton beam and 72 GeV per nucleon with the lead beam.
This is half way between the CM energies of SPS and RHIC, allowing for detailed studies of bottomonium production and
dynamics.

With a nine-month-per-year proton program, one would be able to study the production of quarkonia in 
$pp$, $pd$ and  $pA$ collisions with a statistical accuracy never reached before, especially in 
the target-fragmentation region $x_F \to -1$.  One also has the possibility to explore the region $x>1$ in detail.
High-precision quarkonium-production measurements in $pp$ are of highest relevance to help solving the longstanding puzzles
in $J/\psi$ and $\Upsilon$ production. In particular, it is important to measure observables such as the polarisation
parameters of the direct yields; this means that we should be able to subtract precisely yields from excited states 
such as the $P$-waves through their decay into 
$J/\psi$ and $\Upsilon$ plus a photon. $P$-wave studies are also very important {\it per se}, as well as those of 
the ground states $\eta_c$ and $\eta_b$. A convincing picture of quarkonium production would only be obtained 
once a mechanism or a set of 
mechanisms can be identified as the dominant ones for all quarkonia. Once this is done, most likely also thanks to LHC results, 
quarkonia should be reconsidered as very competitive probes of gluon --and charm quark-- distribution, $g(x)$,  
as it was done 25 years 
ago~\cite{Martin:1987vw,Glover:1987az} -- quarkonium-production yields are proportional to the square of $g(x)$. 
In this case, a specific emphasis on the study of quarkonia
at low $P_T$ and for negative $x_F$ in $pp$ and $pd$ should pose very stringent constraints
on $g_p(x)$ at mid and large $x$ but also on $g_n(x)$ which is pretty much unknown.

As aforementioned, fixed-target experiments provided very important information by accessing the large positive $x_F$ limit, where
novel QCD effects appear and break factorisation. In comparison, the {\it large} negative $x_F$ region, 
which we refer to as the backward limit, has never been explored.
HERA-B is thus far the only experiment which could systematically access the
region of negative $x_F$. It could however not go below $x_F\simeq -0.3$ for $J/\psi$~\cite{Abt:2008ya} for instance.
As we shall elaborate later, the backward limit in $pA$ is interesting in many respects since it is essentially 
different from the forward limit.

The LHC Pb-one-month program will tell us much on Quark-Gluon Plasma (QGP) which should be
created  at $\sqrtS{NN}=72$ GeV in Pb$A$ collisions. Yields close to that of LHC at 5.5 TeV and RHIC at 200 GeV
are expected. These are two orders of magnitude larger than at RHIC at 62 GeV.
The first results from the LHC at 2.76 TeV~\cite{Abelev:2012rv,Chatrchyan:2012np,Aad:2010px} confirmed that the pattern 
of quarkonium --anomalous-- suppression at high energy is very intricate with subtle $y$ and $P_T$ dependences.
Low energy experiments, where recombination process~\cite{Andronic:2003zv} cannot be significant, 
may then play a key role, especially if the recent ultra-granular calorimetry technology
allows one to measure \chic\ and even \chib\ production and suppression in heavy-ion collisions 
--thus far two measurements not available in any other experimental configuration. In such a case, the quest
of the sequential suppression of quarkonia as a QGP thermometer would be realistic again.

A further advantage of a fixed-target set-up is the possibility of polarising the target.
A transverse polarisation would add the possibility 
of studying spin correlations such as the 
non-factorising~\cite{Sivers:1989cc,Brodsky:2002rv,D'Alesio:2007jt,Barone:2010zz} 
aspects of the Sivers 
effect which pins down --for gluon sensitive observables--  
the correlation between the gluon $k_T$ and the nucleon spin. Such measurements could obviously be done
with quarkonia at AFTER, following the pioneer studies by PHENIX~\cite{Adare:2010bd} $p^\uparrow p \to J/\psi X$.

\section{Quarkonium studies with hydrogen target: resolving the production puzzle}

\subsection{Short historical account of the progress since the mid 90's}
Although quarkonia are among the most studied bound-quark systems, 
one should concede that, for the time being, there is no consensus 
concerning which mechanisms are effectively at work in their production in 
high-energy $pp$ collisions, that is at RHIC, at the Tevatron and, recently, at the LHC.  
For recent reviews, the reader is guided 
to~\cite{Lansberg:2006dh,Brambilla:2010cs}
along with some perspectives for the LHC \cite{ConesadelValle:2011fw}.

Historically, the first puzzle was uncovered by the first measurements of the {\it direct}
production of $J/\psi$ and $\psi'$ at $\sqrt{s}=1.8$ 
TeV  by the CDF Collaboration~\cite{Abe:1997jz,Abe:1997yz} whose rates at mid and large $P_T$ were found 
to be much larger than the LO prediction of the QCD-based approach of the 
Colour-Singlet Model (CSM)~\cite{CSM_hadron}. The first NLO evaluations only
appeared  for the $J/\psi$ and $\Upsilon$ for the yields in 2007~\cite{Campbell:2007ws},
 for the polarisation in 2008~\cite{Gong:2008sn}  and  for the $\chi_c$ yields~\cite{Ma:2010vd} in 2010.

In the meantime, other approaches were introduced (\eg~the Colour-Octet Mechanism 
(COM) from NRQCD~\cite{Bodwin:1994jh}) or revived (\eg~the Colour-Evaporation 
Model (CEM)~\cite{CEM}). Unfortunately, these mechanisms --despite their lower predictive
power compared to the CSM-- are also not able to reproduce in 
a consistent way experimental studies of both cross-section and polarisation
measurements for charmonia at the Tevatron \cite{Abulencia:2007us,Affolder:2000nn} along with those 
measured at RHIC~\cite{Adare:2006kf,Abelev:2009qaa,Adare:2009js} and at the 
LHC~\cite{Aaij:2011jh,Aad:2011sp,Aamodt:2011gj,Khachatryan:2010yr,Abelev:2011md}.
As an example, the seemingly solid COM prediction of a transverse polarisation 
of $\psi'$ produced at high $P_T$ is clearly challenged by the experimental measurements~\cite{Abulencia:2007us}.
Recent global fit analyses of $J/\psi$ CO Matrix 
Elements (MEs)~\cite{Butenschoen:2012px,Chao:2012iv}
--consistent with universality--
strongly differ in their conclusion when confronted with polarisation data
 due to a large effect of the choice of the fit sample. This is actually worrisome. 
Such a difficulty of NRQCD can be naturally explained if the charmonium system is too light for the 
relativistic effects to be be neglected\footnote{Along these lines, see the very interesting recent 
study~\cite{Xu:2012am} of the $v^2$ correction for CO channels which seem to be significant.} and  
for the NRQCD~\cite{Bodwin:1994jh} quark-velocity expansion ($v$) to be applied for the  
rather ``light'' $c\bar c$ system. This may very well be so in view 
of the agreement between theory and the available experimental 
data on production in \pp\ of the significantly heavier 
$\Upsilon$. The CSM NLO predictions  including real-emission NNLO contributions at 
$\alpha_S^5$ --the NNLO$^\star$--~\cite{Artoisenet:2008fc} show 
near agreement~\cite{Lansberg:2008gk,Lansberg:2011hi} at mid $P_T$ with the $J/\psi$ and
$\psi'$ data  coming from the Tevatron~\cite{Abe:1997jz,Abe:1997yz,Aaltonen:2009dm} 
and the LHC~\cite{Aaij:2011jh,Aad:2011sp,Aamodt:2011gj,Khachatryan:2010yr}. However, at large $P_T$, 
a gap persists between charmonium data and the CSM predictions. 

As regards the total $P_T$-integrated $\psi$ yield, we should note that
it is also very well described by the sole CS contributions~\cite{Brodsky:2009cf,Lansberg:2010cn}
from RHIC energy all the way to that of the LHC. This is supported by the results of
recent works~\cite{ee} focusing on production at $e^+ e^-$ 
colliders which have posed stringent constraints on the size of $C=+1$ CO contributions which 
can be involved in hadroproduction at low $P_T$; this is
reminiscent of the broad fixed-target
measurement survey of total cross sections~\cite{Maltoni:2006yp} which challenged the universality of
the CO MEs. Finally, let us note that the early claim --based on a LO analysis~\cite{Klasen:2001cu}--
 of evidence for COM in $\gamma \gamma$ $J/\psi$ production has become a difficulty 
of the COM itself at NLO~\cite{Butenschoen:2012px}.

 In the bottomonium sector, relativistic corrections should be smaller and the leading Fock state, that is the CS
$^3S_1$ state, should be dominant. This explains why the inclusion of this sole CS 
channel contribution~\cite{Lansberg:2008gk}  is sufficient to convincingly reproduce the total 
yield~\cite{Brodsky:2009cf,Lansberg:2010cn} -- as for charmonia-- {\it as well as} the cross section differential in 
$P_T$~\cite{Lansberg:2008gk} -- when $P_T^{-4}$ contributions are included-- 
from RHIC~\cite{Abelev:2010am}, the Tevatron~\cite{Acosta:2001gv,Abazov:2005yc} and the
 LHC~\cite{Khachatryan:2010zg,Aad:2011xv,Aaij:2012ve}. 

Finally, it is interesting to mention here the first spin results on quarkonium obtained 
at RHIC. That is the measurement by PHENIX~\cite{Adare:2010bd} of an increasing asymmetry, $A_N$, in 
$pp^\uparrow\to J/\psi X$ for increasing $x_F$, in favour --following the argument 
of~\cite{Yuan:2008vn}-- of the CS dominance of low $P_T$. We will come back to this in the
section~\ref{subsec:SSA}.

\subsection{AFTER-wish-list to solve the puzzles}

Quarkonium measurements in unpolarised $pp$ collisions can be divided in 3 categories: 
(i) yields (differential in $P_T$ or $y$), (ii) polarisation
or spin alignment and (iii) associated production. We postpone the discussion of the last 
topic to section~\ref{subsec:assoc_prod}. For each, we briefly 
discuss available and missing data. At the end, we list the needs for forthcoming analyses, 
most of which would be carried at AFTER.

\subsubsection{Yields and differential cross-section  measurements in high-energy $pp$ 
collisions\protect\footnote{We restrict ourselves to the discussions
of the AFTER, RHIC, Tevatron and LHC energy domain, that is from 100 GeV upwards.}}

In most cases, theoretical predictions are most precise when one refers to {\it direct} yields, those which
do not include any feed-down from higher excited states or $B$ decays --in the case of charmonia. It is therefore
very important to have as many such measurements as possible. Despite this, prompt measurements
may still be useful provided that the feed-down is known in a similar phase-space region.

As for now, direct yields are known --by order of reliability-- for $\Upsilon(3S)$, 
$\psi'$, $J/\psi$, $\Upsilon(2S)$ and $\Upsilon(1S)$. As regards $J/\psi$, the $\chi_c$ 
feed-down measurements have been carried out only twice at colliders for mid $y$ by CDF during 
Run-1~\cite{Abe:1997yz} and by PHENIX~\cite{Adare:2011vq}. The CDF measurements  are
differential in $P_T$ --with only 4 bins, though--, but do not extend to low $P_T$. These analyses have never been cross checked. 
Only a single $\chi_b$ feed-down measurement exist, and is not differential in $P_T$. One cannot tell anything on
a possible variation of the $\chi_b$ feed-down fraction with $P_T$.

The $\chi_c$ and $\chi_b$ feed-down fraction extractions are therefore one of the top priorities in 
quarkonium-production studies. They would provide stringent constraints on the theoretical models. {\it Per se}, 
$P$-wave yields are also very important since they can now also be confronted with accurate theoretical predictions, 
for instance at NLO~\cite{Ma:2010vd} or in $k_T$ factorisation approaches~\cite{Baranov:2011ib,Saleev:2012hi}
(see also section 3.3 of~\cite{Lansberg:2006dh}). Such measurements 
are also of paramount importance at low $P_T$ (below 3 GeV) to test predictions of the total direct yields.

One should also note that extending measurements of $P_T$ spectrum up to very high values --100 GeV or more-- 
is not necessarily optimal, since the dominant production mechanisms in this region are not necessarily
those in the mid-$P_T$ region and because of the presence of large logarithms of $P_T/m_Q$ in the theory at
fixed order in $\alpha_s$.

\subsubsection{Polarisation}

There have been much fewer studies of quarkonium polarisation. For instance, not a single measurement
of the polarisation of the {\it direct}  $J/\psi$, $\Upsilon(1S)$ and $\Upsilon(2S)$ exists or is
even under study. Only such data exist for $\psi'$~\cite{Abulencia:2007us}, although 
with a limited precision, or may be delivered  for $\Upsilon(3S)$. 
Beside this, it has been recently emphasised~\cite{Faccioli:2008dx,Faccioli:2010kd} that the sole 
measurement of the polar --$\theta$-- anisotropy, the so-called
$\alpha$ or $\lambda_\theta$ parameter, is not sufficient to extract all the information about the 
quarkonium polarisation and it can be strongly biased by the acceptance of the experimental set-up. Ideally measurements
should either be done using multiple frames (by changing the spin-quantisation axis) or by studying both the
$\theta$ and $\phi$ angular dependence (2D analysis) of the quarkonium decay products. 

The needs are simple to list --but likely very demanding for LHC experiments for instance. 
One clearly needs a cross check of the prompt --that is direct-- $\psi'$ polarisation in 
2 frames or in 2D, preferably differential in $P_T$ (up to 20 or 30 GeV where theoretical 
predictions do differ) and differential in $y$. The same should also be done for 
$\Upsilon(3S)$. This would be a very good starting point to discriminate models.

However, nothing guarantees that the production mechanisms of these excited states are 
{\it de facto} the same as of $J/\psi$  and $\Upsilon(1S)$. Similar analyses would require 
subtracting the polarisation induced by the $P$-wave feed-down, which is probably 
a highly complicated task. Given the theoretical situation, this is however what seems 
to be required to pin down the specific mechanisms at work in $J/\psi$  and $\Upsilon(1S)$ 
hadroproduction. If this appears to be out of experimental reach, investigating 
associated-production channels will be advantageous. These are discussed in 
section~\ref{subsec:assoc_prod}.

\begin{table}[!hbt]
\centering \setlength{\arrayrulewidth}{.8pt}\renewcommand{\arraystretch}{1.1}
\begin{tabular}{cccc}
 \hline \vspace*{-.15cm} \\
 Target       & $\int\!\! dt{\cal L}$ & ${\cal B}_{\ell\ell}\frac{dN_{J/\psi}}{dy}\Big\vert_{y=0}$&
${\cal B}_{\ell\ell}\frac{dN_{\Upsilon}}{dy}\Big\vert_{y=0}$ \smallskip
\\
\hline \vspace*{-.15cm} \\
100 cm solid H                & 26          & 5.2 10$^8$  & 1.0 10$^6$ \\
100 cm liquid H               & 20            & 4.0 10$^8$  & 8.0 10$^5$ \\
100 cm liquid D               & 24          & 9.6 10$^8$  & 1.9 10$^6$ \\
1 cm Be                      & 0.62         & 1.1 10$^8$  & 2.2 10$^5$ \\
1 cm Cu                      & 0.42         & 5.3 10$^8$  & 1.1 10$^6$ \\
1 cm W                       & 0.31         & 1.1 10$^9$  & 2.3 10$^6$ \\
1 cm Pb                      & 0.16         & 6.7 10$^8$  & 1.3 10$^6$ \\
\multirow{2}{*}{$pp$ 
{\scriptsize low $P_T$ LHC (14 TeV)}
\Big\{ \!\!\!\!} 
                             & 0.05         & 3.6 10$^7$  & 1.8 10$^5$ \\
                             & 2            & 1.4 10$^9$  & 7.2 10$^6$ \\
$p$Pb {\scriptsize  
LHC (8.8 TeV)}               & 10 $^{-4}$    & 1.0 10$^7$  & 7.5 10$^4$ \\
$pp$ {\scriptsize 
RHIC (200 GeV)}              & 1.2 10$^{-2}$ & 4.8 10$^5$  & 1.2 10$^3$ \\
$d$Au {\scriptsize 
RHIC (200 GeV)}              & 1.5 10$^{-4}$ & 2.4 10$^6$  & 5.9 10$^3$ \\
$d$Au {\scriptsize 
RHIC (62 GeV)}               & 3.8 10$^{-6}$ & 1.2 10$^4$  & 1.8 10$^1$ \\
\hline
\end{tabular}
\caption{Nominal yields for \jpsi\ and $\Upsilon$ inclusive production for the quarkonium rapidity $y\in[-0.5,0.5]$  expected per 
LHC year with AFTER with a 7 TeV proton beams on various targets compared to those 
reachable -- also for $y\in[-0.5,0.5]$-- (i) at the LHC    in $pp$
at 14 TeV with the luminosity to be delivered for LHCb and ALICE (which have a low $P_T$ \jpsi\
coverage), (ii) in a typical LHC $p$Pb run at 8.8 TeV, (iii) at RHIC\protect\footnotemark~in $pp$ and (iv) in 
$d$Au collisions at 200 GeV as well as in $d$Au collisions at 62 GeV.
The integrated luminosity is in unit of fb$^{-1}$ per year, the yields are per LHC/RHIC year.  }\label{tab:yields}
\end{table}
\footnotetext{The luminosity for RHIC are taken from the PHENIX decadal plan~\cite{phenix-decadal}.}

\subsubsection{Wish-list summary}

According the above discussion, we can summarise the present needs  as follows:
\begin{enumerate}
\item cross-check  studies of the {\it direct} $J/\psi$ yield ($\chi_c$ only measured once by in $pp$ by CDF and once by PHENIX);
\item cross-check studies of the {\it direct} $\Upsilon(1S,2S)$ yields ($\chi_b$ only measured in $pp$ by CDF (for a single $P_T$ bin));
\item polarisation studies of the {\it direct} yields at least in 2 frames or with a 2D analysis
for \jpsi, \psip~and $\Upsilon(3S)$ (only known for $\psi'$ in 1 frame);
\end{enumerate}

Given the many advantages offered by AFTER, the very large yields --3 orders of magnitude above those at RHIC (see \ct{tab:yields})--, 
a potentially very good acceptance at low $P_T$ thanks to the boost, the likely excellent muon energy resolution,
the availability of novel particle-flow techniques for photon detection in high-multiplicity environment, one is hopeful that
most of --if not all-- these measurements could be carried out thanks to AFTER. Obviously, in 10 years from now, LHC results 
would be also strongly complementing this existing picture with very high energy data.

\section{Mid and large-$x$ gluon-PDF extraction from quarkonium yields}

One of the limitations of Deep-Inelastic-Scattering (DIS) experiments
is that they only directly probe the target quark content. It 
is difficult to directly probe the gluons, although they carry 
a large fraction --$40\%$ at $Q^2\simeq 10$ GeV$^2$-- of the proton momentum. 
Indirect information --mostly at low $x$-- can be obtained from 
the scaling violation of the quark PDFs.
In large-$x$ DIS and DY, PDF extraction is not an easy task  due to the presence of higher-twist
corrections, such as mass effects~\cite{Schienbein:2007gr} and direct processes~\cite{Berger:1979du}. 
In this region, sum rules are also practically useless since the PDFs are strongly suppressed
for $x\to 1$. As a consequence, the gluon distribution is very badly known for 
$x>0.2$ at any scale, see for instance Fig 1. of~\cite{Brodsky:2012vg}.

In this context, the information offered by quarkonium production 
in the target region at AFTER should be invaluable. In the conventional approach, quarkonia are 
produced at high energy by gluon fusion at scales of the order of their mass, 
thence large enough to use perturbative QCD (pQCD). 
As discussed above, a number of puzzles are complicating the picture. The situation has changed since 
the pioneering analyses of gluon 
PDF extraction with quarkonium data~\cite{Martin:1987vw,Glover:1987az} and, nowadays, few people still consider
quarkonia as reliable probes of gluon distribution. High-$P_T$ jet or even prompt photon studies
are preferred. We are however very confident that these puzzles would be solved by the systematic high-precision
studies mentioned above in addition with the forthcoming LHC results at high energies and most likely as well
for higher $P_T$. Let us in addition mention the possibility
of a contribution at large $x$ by IC via $gc\to J/\psi c X$~\cite{Brodsky:2009cf} or 
via diffractive reaction~\cite{Brodsky:2006wb}. These are 
discussed in section~\ref{sec:heavy_quarks}. Like for other hadronic reactions
where different PDFs are involved, extraction of one specific PDF, once the other contributions are known, 
is possible; this thus calls for the extraction of $c(x)$ and to a lesser extent to $b(x)$.

That being said, the use of $C=+1$ quarkonia, in particular the $\eta_{c,b}$ 
and $\chi_{c,b}(^3P_2)$,  should in any case be more reliable than of $J/\psi$.  $\eta_{c,b}$ 
and $\chi_{c,b}(^3P_2)$ are produced at LO without 
a final-state gluon~\cite{Lansberg:2006dh,Brambilla:2010cs,Maltoni:2004hv}, hence (i) with very competitive rates and  
(ii) via a Drell-Yan like kinematics for which the gluon momentum fractions are simply related to the quarkonium 
rapidity. Last but not least, large QCD corrections to the $P_T$ spectrum predictions,
as seen for $\psi$ and $\Upsilon$~\cite{Campbell:2007ws,Artoisenet:2007xi,Gong:2008sn,Artoisenet:2008fc}, are not expected
since the leading-$P_T$ scaling is already reached at NLO.

A modern ultra-granular electromagnetic calorimeter~\cite{Thomson:2009rp} should allow one to
 study both $\chi_{c,b}(^3P_2)$ through $\ell^+ \ell^- \gamma$ decays and $\eta_{c}$ in the $\gamma\gamma$ 
channels down small $P_T$ and to large negative $x_F$. With a good Particle IDentification (PID), the study of the $p\bar p$ 
decay channel (see \eg~\cite{Barsuk:2012ic}) is reachable, opening the door to a systematic study of
all the hidden charm resonances. Doing so, it is reasonable to consider that, with charmonia, $g(x)$ could be measured
accurately for $Q^2\simeq 10$ GeV$^2$ from $x$ as low as $10^{-3}$, and with bottomonia,  
for $Q^2\simeq 100$ GeV$^2$ from $x=3 \times 10^{-2}$ up to $x\simeq 1$ in the range $y_{onium} \in [-4.8,1]$.
 The case for bottomonium is certainly stronger since (i) its production certainly lies in the perturbative domain
of QCD for any $P_T$, (ii) higher-twist contributions and relativistic corrections should also be small, 
(iii) CO contributions can certainly be neglected, and (iv) the impact of IB should be ten times smaller than that of IC (see section~\ref{sec:heavy_quarks}).

\section{Quarkonium production in proton-deuteron collisions: gluons in the neutron}

The most competitive way to study the partonic structure of the neutron
is to use combined measurements with hydrogen and deuterium targets.
Studies of the gluon distribution in the neutron, $g_n(x)$, are singularly more complicated
than those of quarks based on DIS experiments (see~\cite{Airapetian:2011nu} 
for a recent account of existing results) and DY (see \eg~\cite{Baldit:1994jk}).
An isospin asymmetry of the sea quarks has for instance been uncovered~\cite{Baldit:1994jk}.

As regards the gluon in the neutron, quarkonium hadroproduction seems to be an ideal probe. 
Muo-production $J/\psi$ studies on proton and deuterium targets by the NMC collaboration~\cite{Allasia:1990zx} 
showed that --within their 15\% uncertainty--  $g_n(x)$ was similar to $g_p(x)$. 
The E866 $\Upsilon$ analysis~\cite{Zhu:2007mja} in $pp$ and $pd$  confirmed that 
$g_n(x,Q^2\simeq 100 \hbox{ GeV}^2)\simeq g_p(x,Q^2\simeq 100 \hbox{ GeV}^2)$ for $0.1\leq x \leq 0.23$. 
Unfortunately, this measurement could not be done simultaneously for $J/\psi$ which would have probed
$g_n(x)$ at lower $Q^2$ and lower $x$.

Using a 1m-long  deuterium target, one would obtain (see \ct{tab:yields}) $10^9$ \jpsi\ and
$10^6$ $\Upsilon$ decaying in muon pairs in one unit of $y$ in proton-deuteron collisions. Such high-precision 
measurements may allow for the discovery of a difference between $g_n(x)$ and $g_p(x)$. In any case, such an analysis
would allow for the extraction of $g_n(x)$ in a significantly wider $x$ range and at lower $Q^2$ with $J/\psi$. 
It is important to keep in mind that one does not anticipate other facilities where $pd$ collisions could be studied 
in the next decade -- at least at high enough luminosities and energies where such measurements could be carried out.

\section{Quarkonium production in $pA$ collisions: taming the nuclear-matter effects}
\label{sec:nuclear_matter}

The study of hard hadronic processes in $pA$ collisions gives the opportunity to study
a large number of very interesting QCD effects, among others
\begin{itemize}
\item the modification of the partonic densities inside bound nucleons;
\item the propagation of the hadrons in the nuclear matter during their formation;
\item the energy loss of partons which is induced by the nuclear matter, be it absolute or fractional;
\item the colour filtering of IC in the nuclear matter;
\end{itemize}

  A high luminosity fixed-target experiment
with versatile target choice is the best set-up to explore this physics, which
is of significant importance for interpreting the physics of hard scatterings 
in $AA$ collisions discussed in section~\ref{sec:AA}. This physics is 
genuinely at the small-distance interface between particle and nuclear physics, it
deserves thus much efforts which could be very rewarding in terms of discovery 
of novel QCD aspects.

It is clear that AFTER is the
ideal experiment to illuminate this physics in the sub-TeV domain, taking into account
the strength and the weakness of SPS, Fermilab and RHIC experiments.
For instance, the major disadvantage of studying (proton-)ion collisions with collider experiments 
is the intrinsic difficulty to change the beam species. During 
the past ten years, RHIC studied four kinds of collisions: $pp$, $d$Au (equivalent 
to $p$Au), CuCu and AuAu. The luminosities are also severely limited. 
No $\psi'$ yield has been measured so far in $d$Au collisions, while the fixed-target experiment 
E866 with a versatile choice of targets had enough luminosity to find a difference of the absorption between $J/\psi$ and $\psi'$ 
at low $x_F$~\cite{Leitch:1999ea}. However, E866 could not take data simultaneously
for charmonium and bottomonium, neither could it look at $P$ wave as did Hera-B. On the other hand, 
SPS experiments were limited in their rapidity coverage, whereas PHENIX at RHIC can give information over
more than 4 units of rapidity. The limitation in energy of SPS
also severely limited $\Upsilon$ studies.

To avoid the limitations of past
experiments, 
it appears that the capabilities of a new experiment should at least be:
\begin{itemize}
\item Collection of large statistical samples;
\item Wide coverage of rapidity and $x_F$;
\item Wide coverage of $A$ ;
\item Simultaneous measurements of most of the quarkonium and open heavy-flavour states. 
\end{itemize}

This is precisely what AFTER can offer above 100 GeV with a multi purpose detector and a large coverage from
$y_{cms}\simeq 0$ down to the target rapidity\footnote{An idea similar to AFTER has been proposed in~\cite{Kurepin:2011zz}
using a ribbon-like lead target at the interaction point of the ALICE detector. This also offers rates higher than at RHIC but does
not have the versatility of an extracted beam line. In particular, it does not allow for $pp$ and $pd$ 
measurement with long hydrogen and deuterium targets as discussed above. Neither does it allow for polarised-target analyses.}. 
In practice, in the far  -- and less far-- negative $x_F$ region,
 we would like to master the effect from the gluon nPDF and from the survival probability of the quarkonium
 along its way out of the nucleus. These effects should significantly affect quarkonium production, 
implying in return that they can be analysed thanks to quarkonium. 

The plan for high-precision $pA$ quarkonium-production studies essentially bears
on the very large yields detailed in~\ct{tab:yields}
at AFTER in $p$Pb collisions for instance, $10^9$ $J/\psi$ and $10^6$ $\Upsilon$ 
per year and per unit of rapidity. The  target versatility would of course be a strong asset
to investigate the dependence on the impact-parameter, $\vect b$, of nuclear-matter effects, 
in particular that of the nPDFs~\cite{Klein:2003dj}. The precision and the interpretation of the RHIC studies 
--using the sole $d$Au system-- is indeed limited by the understanding and the measurements 
of the so-called  centrality classes. A further key requirement here is 
to study multiple states to get a handle on factorisation breaking
effects in this particular phase-space region.

Let us also note that low-$x$ gluon determination is one of the 
main aims of Electron-Ion Collider projects (eRHIC, ELIC, LHeC), principally by very precise studies of DIS on proton 
 and nucleus targets as well as of photo- and electro-production of $J/\psi$. Both EIC and AFTER projects are 
essentially complementary. However, it is worth noting that gluon-nPDF extraction via quarkonium 
studies in $ep$ or $eA$ collisions will also require progress
in understanding quarkonium hadroproduction. It is reasonable to say
that a reliable extraction of $g(x)$ from $J/\psi$ cross sections in $ep$ and $eN$ collisions 
can only be achieved once the hadroproduction puzzles are behind us.

\subsection{Quarkonium production as probe of gluon nPDFs: shadowing, antishadowing and EMC effect}

Nuclear PDFs (nPDFs) contain a wealth of information about the dynamics of parton
in nuclei. For $x$ larger than unity, they even contain information about the correlations
between the nucleons within the nuclei; these are pretty unknown at the GeV scale.
Fermi motion is known to alter PDFs at very large $x$ ($<1$).
Then, a depletion of the PDFs is observed, for $0.3 \leq  x \leq 0.7$. 
It is known since 1983, as the EMC effect, but there is no consensus on its physical origin.
At lower $x$, one observes an excess of partons compared to free nucleons at mid $x$ -- 
the antishadowing. This observation was made in electron-nucleus deep-inelastic reactions, 
but appears to be absent in the case of Drell-Yan processes in $pA$ 
and neutrino charged-current reactions~\cite{Kovarik:2010uv}. It could be 
 that antishadowing is quark or anti-quark specific because of the flavour dependence 
of Regge exchange in the diffractive physics underlying Glauber 
scattering~\cite{Brodsky:2004qa,Brodsky:1989qz} or because it is
a higher twist effect. At small $x$, below say 0.05,  a further depletion of 
nPDFs --the nuclear shadowing~\cite{Glauber:1955qq,Gribov:1968jf}-- is observed.

Whereas these effects can be studied for quark distributions, there are usually very
difficult to probe in the gluon sector. Measurement of low-$x$ gluon shadowing is 
for instance one of the EIC flagships. 
At mid- and large-$x$, it is important to note that gluon-nPDF fits are plagued by large
--and sometimes fit-dependent-- uncertainties. The amount of the EMC suppression is 
actually pretty much unknown~\cite{Eskola:2009uj}, except for a loose constraint set by 
momentum conservation. Quarkonia can be key players here. RHIC experiments were the first 
to extend quarkonium studies in $p(d)$Au collisions above the 100 GeV limit hinting at 
gluons shadowing in the Au nucleus~\cite{Adler:2005ph}. RHIC $\Upsilon$-production data also 
hint~\cite{Ferreiro:2011xy} at a gluon EMC effect stronger than for quarks, but higher 
precision data are needed. It will be difficult for PHENIX and STAR to provide definitive data.

In this context, the large charmonium and bottomonium yield we expect at
$\sqrt{s_{NN}}=115$ GeV, $10^9$ $J/\psi$ and $10^6$ $\Upsilon$ per year and per unit of rapidity 
(see \ct{tab:yields}), should allow for high-precision  $pA$-production studies and in turn
give us confidence in gluon-nPDF extraction with quarkonia. 
A good enough resolution would allow the measurement of ratios of yields such
as $N_{J/\psi}/N_{\psi'}$. We would have access to open charm and beauty with vertexing. Other
ratios such as  $N_{J/\psi}/N_D$ and $N_{\Upsilon}/N_B$ --where the nPDF effect may cancel-- could then be extracted. 
Good photon calorimetry would make a systematic study of $\chi_c$ and $\chi_b$ possible, extending 
the measurements of HERA-B \cite{Abt:2008ed}. AFTER could provide the first study of $\eta_c$ 
inclusive production in $pA$ collisions. A combined analysis of these observables would 
certainly put stringent constraints on the gluon distribution in nuclei at mid and large $x$, given that 
they would also help understand other effects at work on which we elaborate now.

\subsection{Additional physics in quarkonium production in $pA$}

In the negative $x_F$ region, the mesons are also fully formed when escaping the nucleus
with a survival probability which is minimal and related to their physical size. 
In more forward configurations, where the $Q\bar Q$ is boosted in the nucleus rest frame, 
the survival probability is not necessarily related to their size;  one thus often 
parametrises it in terms of an effective cross section. 
This picture could be checked in detail with a careful study of ratios of yields
of different quarkonia (see \eg~\cite{Koudela:2003yd}). 
A scan in $x_F$, thus in $\sqrt{s_{\psi N}}$, would help us understanding
the physics underlying this effective break-up cross section,
which may reveal higher twist effects~\cite{Kopeliovich:2001ee}.
{\it A priori}, this region is not affected by fractional energy loss~\cite{Arleo:2010rb}, neither by colour
filtering of IC~\cite{Hoyer:1990us}, relevant at large $x_1$, not at large $x_2$.  In the negative $x_F$ region, 
IC may be a natural source of charmonium production but it would not show the $A^{2/3}$ suppression
discussed in section~\ref{sec:heavy_quarks}.

All of these aspects can also be investigated with DY, prompt photon, photon-jet correlation 
and heavy-flavour measurements
at AFTER. This would cross-check --or feed information in-- the interpretation done with quarkonia.
We emphasise once again that HERA-B is so far the only experiment which could easily access the
region of negative $x_F$, down to -0.3 for $J/\psi$~\cite{Abt:2008ya}. No other facility could ever go  below that.

\section{Heavy-quark (n)PDFs and quarkonia}
\label{sec:heavy_quarks}
From the non-Abelian QCD couplings of the gluons~\cite{Brodsky:1984nx,Franz:2000ee}, one expects
the probability of the intrinsic Fock state in a proton $|uudQ\bar Q\rangle$ to fall 
as $1/M^2_{Q \bar Q}$. The relevant matrix element is the cube of the QCD field strength $G^3_{\mu \nu}$,  
in contrast to QED where the relevant operator is $F^4_{\mu \nu}$; the probability of intrinsic 
heavy leptons in an atomic state is suppressed as $1/m^4_\ell.$   
It can be shown that the heavy-quark pair $Q \bar Q$ in the intrinsic Fock state  is then primarily a colour-octet. 
The ratio of IC to IB also scales as $m_c^2/m_b^2 \simeq 1/10$.
Many aspects of this physics can be studied at AFTER with quarkonia both in $pp$ and $pA$ collisions.

\subsection{Intrinsic charm and beauty in the proton}
\label{subsec:IC_proton}

Initial parametrisations of the charm and bottom quark PDFs used in global fits of the proton structure 
functions only have support at low $x$ since one usually assumes that they only arise from gluon splitting 
$g \to Q \bar Q$ from $Q^2$ above $4m_Q^2$ and
that the IC or IB component is negligible.  
Inaccurate predictions can result from this assumption, especially in large $x_F$ or $x_T$ heavy-hadron production.
In agreement with the EMC measurements~\cite{Harris:1995jx}, IC predicts that the charm structure function 
has support at large $x$ in
excess of DGLAP extrapolations~\cite{Brodsky:1980pb}.  It is surprising that the original 1983 EMC experiment 
which first observed a large signal for charm at 
large $x$ in $\gamma^\star p \to c X$  has never been repeated.
Lai, Tung, and Pumplin~\cite{Pumplin:2007wg} emphasised that $c(x)$ may have been underestimated in usual global fit, but
dedicated measurements are still awaited for.

Careful analyses of the rapidity distribution of open- or hidden-charm hadrons in a 
fixed-target set-up at $\sqrt{s}=115$ GeV are 
be therefore very important --especially at backward rapidities-- to learn more on 
these aspects of QCD. In particular, it has been shown~\cite{Brodsky:2009cf} that, at $\sqrt{s}=200$ GeV, a significant fraction 
of the $J/\psi$ yield is expected to be produced in association with a charm quark. It was also emphasised that the 
measurement of the rapidity dependence of such a yield would provide a complementary handle on $c(x)$. 
Such a measurement would efficiently be done by triggering on $J/\psi$ events then by looking for
$D$ {into its $K\pi$ decay for instance}. At large $|x_F|$, diffractive $J/\psi$ 
production~\cite{Brodsky:2006wb} should arise from IC coalescence.
Quantitative predictions of the cross section are however lacking and much can still be learnt.
An excess of double $J/\psi$ events may also sign the presence of IC. 
We believe that a set of precise measurements
as the one mentioned above would certainly help in probing IC and measuring its size.

\subsection{Intrinsic charm and beauty in the nucleus}

AFTER is unique for its access to the negative $x_F$ region, where IC and IB component, 
not in the projectile, but in the target can have an important effect. As 
aforementioned, the IC Fock state has a dominant colour-octet 
structure: $\vert(uud)_{8C} (c \bar c)_{8C}\rangle$. In $pA$ collisions for $x_F\to 1$,  the colour 
octet $c \bar c$ comes from the projectile and bleach into a colour singlet by gluon exchange 
on the front surface of a nuclear target. It then coalesces to a $J/\psi$ which interacts 
weakly through the nuclear volume~\cite{Brodsky:1989ex}. An  $A^{2/3}$ dependence of the rate
is expected, which corresponds to the  front-surface area. This combines with the usual pQCD $A^1$ contribution
at small $x$. 
This combination is actually consistent with charmonium production 
observed by the CERN-NA3 ~\cite{Badier:1981ci} and the Fermilab E866 collaborations~\cite{Leitch:1999ea}.
Because of these two components, the cross section violates perturbative QCD factorisation for hard 
inclusive reactions~\cite{Hoyer:1990us}. Other factorisation-breaking effects exist such
as Sudakov suppression induced by the reduced phase space for gluon radiation at large $x_F$~\cite{Kopeliovich:2005ym},
fractional energy loss~\cite{Arleo:2010rb}, etc. As we discussed in section~\ref{sec:nuclear_matter}, they all deserve careful analyses.

{For negative $x_F$, the IC emerges from the nucleus and can be affected by
 nuclear modifications such as anti-shadowing, EMC or Fermi motion. One does not
expect colour filtering anymore. AFTER provides a unique opportunity to check this prediction.
It has also to be noted that in Pb$p$ collisions at 72 GeV, large negative $x_F$
charmonium production would be become again sensitive to colour filtering of IC. }

\section{Quark-gluon plasma and the quarkonium sequential suppression}
\label{sec:AA}

Charmonium suppression in relativistic heavy-ion collisions has been first proposed 
by Matsui and Satz in 1986~\cite{Matsui:1986dk} as a probe of the formation 
of a Quark Gluon Plasma (QGP) . Since then,  many experiments have provided
important results on \jpsi\ production in $AA$ collisions for instance at
the CERN-SPS at \sqrtS{NN}=17 GeV~\cite{Abreu:1997jh,Abreu:1999qw,Abreu:2000ni} and 
at BNL-RHIC at \sqrtS{NN}=200 GeV~\cite{Adare:2006ns,Adare:2008sh,Tang:2011kr}. Results starts now to 
flow in from the CERN-LHC at \sqrtS{NN}=2.76 TeV~\cite{Abelev:2012rv,Chatrchyan:2012np,Aad:2010px}.
These results tend to indicate that the \jpsi-production 
cross section is indeed modified by the Hot and Dense Matter (HDM) produced at SPS and RHIC, as 
it was predicted almost 25 years ago. However, a definite 
and precise description of the HDM effects on \jpsi\ production is not still at reach.

Among the arguments that can be raised to explain the situation, the lack of knowledge of the nuclear-matter 
effects is the most relevant one; we discussed it in section~\ref{sec:nuclear_matter}. 
One can also argue on the fact that, so far, the main 
experimental results concern \jpsi\ production only. Even though \psip\ production has been studied 
in $AA$ at SPS, the statistics available are poor; these are (almost) absent at RHIC. This
precludes any sensible interpretation. Other quarkonium measurements such as those of \chic\ are so difficult to 
perform that the available data are useless for our purpose.
Such a measurement is however of fundamental  interest since around 30$\%$ of the produced \jpsi\ 
come from the \chic\ decay.  \chic\ and \psip\ may melt at a lower QGP temperature than the \jpsi, 
hence a sequential melting pattern. Finally, the  statistics of \upsi\ measured at SPS
is too low to draw any firm conclusion and, at RHIC, the situation is roughly the same.

The LHC experiments will provide detailed results on $AA$ collisions at an energy 
--\sqrtS{NN}=2.76 and 5.5 TeV-- never reached before and will offer the possibility to study both charmonium 
(\jpsi\ and \psip) and bottomonium ($\Upsilon(nS)$) states, shedding light on 
quarkonium behaviour in such a new HDM regime where most of them are expected to be melted. 
The process of $c\bar c$ recombination~\cite{Andronic:2003zv} might come into play at such high energies. 
The first results on \jpsi\ at 2.76 TeV~\cite{Abelev:2012rv,Chatrchyan:2012np,Aad:2010px} show
an unexpected pattern, difficult to explain.
In addition, CMS~\cite{Chatrchyan:2011pe} has observed a suppression of excited 
$\Upsilon(2S)$ and $\Upsilon(3S)$ states relative to $\Upsilon(1S)$ in these PbPb collisions.
Further data will complete this list, whose interpretation will remain complicated by the
lack baseline of $p$Pb collisions at the same energy and in the same CM rapidity coverage, even though $p$Pb runs are planned.

Low energy experiments, where recombination processes~\cite{Andronic:2003zv} cannot be significant, 
should be very complementary to the forthcoming LHC results. Provided that excited states and 
different nucleus-nucleus $AB$ systems can also
be studied, it is reasonable to argue that the sequential suppression pattern of quarkonium
could be observed and used as a thermometer of the QGP in central $AA$ collisions. It is also
important to have a good control of nuclear-matter effects which, as discussed above, would then be studied
at 115 GeV in $pA$ collisions and which can also be studied at 72 GeV in Pb$p$ collisions with an hydrogen 
target.

\begin{table}[!hbt]
\centering \setlength{\arrayrulewidth}{.8pt}\renewcommand{\arraystretch}{1.1}
\begin{tabular}{cccc}
\hline \vspace*{-.15cm} \\
Target       & $\int\!\! dt{\cal L}$ & $\left.{\cal B}_{\ell\ell}\frac{dN_{J/\psi}}{dy}\right\vert_{y=0}$&
$\left.{\cal B}_{\ell\ell}\frac{dN_{\Upsilon}}{dy}\right\vert_{y=0}$
\\\vspace*{-.15cm}\\
\hline \vspace*{-.15cm} \\
100 cm solid H       & 1100   & 4.3 10$^6$   & 8.9 10$^3$ \\
100 cm liquid H      & 830   & 3.4 10$^6$   & 6.9 10$^3$ \\
100 cm liquid D      & 1000   & 8.0 10$^6$   & 1.6 10$^4$ \\
1 cm Be             & 25    & 9.1 10$^5$   & 1.9 10$^3$ \\
1 cm Cu             & 17    & 4.3 10$^6$   & 0.9 10$^3$ \\
1 cm W              & 13    & 9.7 10$^6$   & 1.9 10$^4$ \\
1 cm Pb             & 7    & 5.7 10$^6$   & 1.1 10$^4$ \\
$d$Au {\scriptsize 
RHIC (200 GeV)}     & 150   & 2.4 10$^6$   & 5.9 10$^3$ \\
$d$Au {\scriptsize 
RHIC (62 GeV)  }    & 3.8   & 1.2 10$^4$   & 1.8 10$^1$ \\
AuAu {\scriptsize 
RHIC (200 GeV)}     & 2.8   & 4.4 10$^6$   & 1.1 10$^4$ \\
AuAu {\scriptsize 
RHIC (62 GeV)}      & 0.13  & 4.0 10$^4$   & 6.1 10$^1$ \\
$p$Pb {\scriptsize 
LHC (8.8 TeV) }     & 100   & 1.0 10$^7$   & 7.5 10$^4$ \\ 
PbPb {\scriptsize 
LHC (5.5 TeV) }     & 0.5   & 7.3 10$^6$   & 3.6 10$^4$  \\
\hline
\end{tabular}
\caption{Nominal yields for \jpsi\ and $\Upsilon$ inclusive production for $y\in[-0.5,0.5]$ 
expected per LHC year with AFTER
with a 2.76 TeV lead beam on various targets compared to the projected nominal yield (i) in Pb$p$ and PbPb runs
of the LHC at 8.8 and 5.5 TeV as well as (ii) in $d$Au and AuAu collisions at 200 GeV and 62 GeV at RHIC. 
The integrated luminosity is in unit of nb$^{-1}$ per year, the yields 
are per LHC/RHIC year.  }\label{tab:yieldsPb}
\end{table}

\ct{tab:yieldsPb} displays the expected \jpsi\ and $\Upsilon$ yields with
the 2.76 TeV Pb beam on several targets. They are compared to those expected per year at 
RHIC in $d$Au and AuAu (at $\sqrtS{NN}=62$ and 200 GeV), at the LHC in Pb$p$ (at \sqrtS{NN}$=8.8$ TeV) 
and  in PbPb (at $\sqrtS{NN} =5.5$ TeV) at their nominal luminosity. 
The yields in PbPb (at $\sqrtS{NN}=72$ GeV) are similar to 
(100 times larger than) those expected in a year at RHIC for AuAu at $\sqrtS{NN}=200$ GeV 
(62 GeV) and also similar to that to be obtained 
during one LHC PbPb run. This is remarkable considering  the lower cross section at AFTER 
because of the lower energies. We note that the same ratios also apply for 
the other quarkonium states. The baseline for such collisions could be measured with the Pb beam  
with a 100cm thick H target, with very competitive rates.

Thanks to recent developments in ultra-granular calorimetry techniques, one expects be able to study other
 charmonium states such as \chic\  in its $\jpsi+\gamma$ decay channel.  This
would help towards the understanding of quarkonium anomalous suppression, especially given
the likely absence of recombination process~\cite{Andronic:2003zv} at this energy. Thanks to the large 
\jpsi\ sample, polarisation studies could be carried out providing useful complementary information~
\cite{Gupta:1998ut,Ioffe:2003rd,Faccioli:2012kp}.

\section{New observables in quarkonium physics at $ \sqrt{s}=115$ GeV}

\subsection{Quarkonium single transverse spin asymmetry}
\label{subsec:SSA}
Target polarisation (see \eg~\cite{Crabb:1997cy}) is an essential benefit
of fixed-target experiments. A transverse polarisation for instance 
allows for Single Spin Asymmetry (SSA) measurements in production reactions.
These SSAs in hard reactions give a handle on a novel class of parton distribution functions, known as 
“Sivers functions”~\cite{Sivers:1989cc} (see~\cite{D'Alesio:2007jt,Barone:2010zz} for recent reviews). 
These functions express  a correlation between 
the transverse momentum of a parton inside the proton, and the proton-spin vector. As such 
they contain information on orbital motion of partons in the proton. 
Nearly nothing is known about gluon Sivers functions which can be probed with quarkonia.
These SSA are believed to be due to the rescattering of the quarks and gluons in the 
hard-scattering reactions, and in general 
they do not factorise in the standard pattern expected in perturbative QCD. 

Recently, measurements from PHENIX~\cite{Adare:2010bd} have shown that the transverse 
SSA in $p^\uparrow p \to J/\psi X$ deviates significantly from zero at $x_F\simeq 0.1$, 
\ie\footnote{$x^\uparrow_p$ is the momentum fraction of the parton emerging from the 
polarised proton.}~$x^\uparrow_p\simeq 0.1$ . According to the analysis of~\cite{Yuan:2008vn}, this hints at a 
dominance of a colour-singlet mechanism at low $P_T$ and at a non-zero gluon Sivers effect. Recently,
$\jpsi$ and $\Upsilon$ SSA studies at AFTER have be investigated~\cite{Liu:2012vn} with the hypothesis
that their production is initiated by $q\bar q$ fusion.
AFTER at $\sqrt{s}=115$ GeV with a high luminosity and
 a good coverage in the rapidity region of the transversally polarised-target  (mid and large 
$x_p^\uparrow$), may be extremely competitive and complementary to the other existing high-energy 
particle physics spin projects, in particular as regards gluon Sivers functions. These would be studied
via SSA most likely for all the states for which yield measurements are possible and up to so far unexplored
$x^\uparrow_p$.

\subsection{Associated-production channels}
\label{subsec:assoc_prod}

\subsubsection{Quarkonium-jet/hadron correlation}

Quarkonium-jet/hadron azimuthal correlation analysis is a simple measurement which does not require  
large statistical  sample -- only an azimuthal distribution is involved-- and is not very demanding in experimental 
requirements. In the nineties, UA1 confronted their  distributions of charged tracks against 
Monte Carlo simulations for a $J/\psi$ coming 
from a $B$ and  a $J/\psi$ coming from a 
$\chi_c$~\cite{Albajar:1987ke,Albajar:1990hf}. The idea was that the activity around
the $J/\psi$ would be higher in non-prompt events (because of the remnants of the $B$ decay into \jpsi) 
than in prompt events, which were thought to come dominantly from $\chi_c$ which can be produced without
nearby gluons. The same technique has been used by STAR with LO Pythia
outputs~\cite{Abelev:2009qaa}.

At present time, we however expect  more complex distributions
even for the prompt yield, be it dominated by CO or CS transitions -- 
$J/\psi$+ 2/3 gluons process may indeed be significant. 
 It is therefore not clear whether such  analyses
are suitable to evaluate the $B$-feed-down. Yet, they are rather easy
to implement and may offer interesting complementary results to
mere yield measurements.  STAR is now working on the analysis of 
$\Upsilon$ data~\cite{Cervantes:2011zz}. Such measurements can be carried out at any
set-up with a good azimuthal coverage. This should be the case of AFTER.

\subsubsection{Double-quarkonium production}

Next on the list is double quarkonium production. $J/\psi$ pairs have already been
observed at mid and large $x_F$ by NA3 30 years ago~\cite{Badier:1982ae} !
These results were consistent with double-IC Fock states~\cite{Vogt:1995tf}.
 The first analysis at the LHC removing the $B$ feed-down has been released 
by LHCb~\cite{Aaij:2011yc}. Note however that it is at significantly smaller 
$\langle x_F\rangle \simeq 4 \times 10^{-2}$ despite the forward rapidities accessed by LHCb. 
Such measurements can be done at AFTER down to large
negative $x_F$. Double-$\Upsilon$ production could also be looked for. 
Kinematical distributions are also worth being investigated. To learn more on these processes, 
it may be interesting to look at $J/\psi+\psi'$ and $J/\psi+\chi_c$ associated
production. NLO theoretical predictions are however lacking for the time being.

\subsubsection{Quarkonia plus open heavy flavour}

As discussed in section~\ref{subsec:IC_proton}, the measurement of the rapidity dependence of
the cross section of associate production of $\jpsi + D$ is an interesting probe of 
IC~\cite{Brodsky:2009cf}. This process can actually be a significant source of inclusive production of
charmonia. Its study is therefore very important both at low and large $P_T$. 
For instance $pp\to J/\psi + c\bar c$ is the dominant contribution in the CSM at 
$\alpha_s^4$~\cite{Artoisenet:2007xi} at large $P_T$. Prompt $J/\psi+b$ may also be an interesting
probe~\cite{Gang:2012js} provided that the dominant background from non-prompt $J/\psi + b$
can be efficiently removed. This maybe done by looking at pairs of one prompt $J/\psi$  and
one non-prompt $J/\psi$.

The measurement of the dependence on $J/\psi+D$ invariant-mass yield or on $\Delta \phi$ may also provide
important information~\cite{Brodsky:2009cf,Berezhnoy:2012xq} about the novel phenomena, \eg~ colour transfers beyond NRQCD
factorisation~\cite{Nayak:2007mb}. The measurement of the polarisation of the $J/\psi$ 
produced with a $D$ meson~\cite{Lansberg:2010vq} may also be a discriminant between production models.
All these measurements can be efficiently performed at AFTER.

\subsubsection{Quarkonium plus isolated photon}

The production of the $\jpsi$ and $\Upsilon$ in association with a prompt/isolated $\gamma$ is very likely
a useful probe to feed in CO ME global fits since CO fragmentation contributions
are sensitive to the $C=+1$ CO transitions, whereas in the inclusive case, it
is rather the $C=-1$, $^3S_1^{[8]}$, transition which is involved in $g^\star \to J/\psi X$.
Colour singlet contributions are also naturally larger.
 
In a sense, this process is the continuum background of the resonant $\chi_c\to J/\psi +\gamma$ signal.
However, we are interested in the region where the photon can be easily detected, \ie~ 
at large enough $P_T$ and isolated from other charged hadrons. The  
invariant mass of the $J/\psi-\gamma$ pair is not fixed.
The direct cross section and polarisation in the CSM have been evaluated at the LHC energy at NLO~\cite{Li:2008ym} and at 
NNLO$^\star$~\cite{Lansberg:2009db} accuracy.

\section{Conclusions}

We have  discussed the numerous contributions that A Fixed Target ExpeRiment on the LHC beams, AFTER,
can provide to advance of quarkonium physics. We expect that it can be, in conjunction with the LHC, the
key experiment to help us put an end to quarkonium-production controversies in $pp$ collisions.
In $pd$ collisions, quarkonium measurements at AFTER may be the very first to discover
that gluon distribution in the neutron is not necessarily equal to that in the proton.
In $pA$ collisions, AFTER can produce the largest quarkonium yields ever observed. Its unique access to the far
backward region, where novel QCD effects may be at work in nuclear matter, can be a decisive advantage. 
In Pb$A$, AFTER can expediently
complement SPS, RHIC and LHC results in an energy range unexplored so far. This adds to
a very good potential for quarkonium excited-state studies in hot and dense nuclear matter.
Finally, in polarised $pp$ collisions, precision studies of single transverse spin asymmetries in 
quarkonium production at AFTER can be essential in the understanding of the gluon Sivers effects, thus of
the gluon-motion contribution to the proton spin.

\section*{Acknowledgements}

We wish Joseph Cugnon and Hans-J\"urgen Pirner continuous success and a fulfilling second career.

This research was supported in part by the Department of Energy, contract DE--AC02--76SF00515 and by
the France-Stanford Center for Interdisciplinary Studies (FSCIS).

\end{document}